\documentclass{vgtc}                          

\ifpdf
  \pdfoutput=1\relax                   
  \usepackage{graphicx}                
  \DeclareGraphicsExtensions{.pdf,.png,.jpg,.jpeg} 
\else
  \usepackage{graphicx}                
  \DeclareGraphicsExtensions{.eps}     
\fi%

\graphicspath{{figures/}{pictures/}{images/}{./}} 

\usepackage{microtype}                 
\PassOptionsToPackage{warn}{textcomp}  
\usepackage{textcomp}                  
\usepackage{mathptmx}                  
\usepackage{times}                     
\usepackage{cite}                      
\usepackage{tabu}                      
\usepackage{booktabs}                  
\usepackage{csquotes}

\onlineid{1055}

\vgtccategory{Empirical study}

\title{Beyond Visuals : Examining the Experiences of Geoscience Professionals With Vision Disabilities in Accessing Data Visualizations}

\author{Nihanth W. Cherukuru \thanks{e-mail: ncheruku@ucar.edu} %
\and David A. Bailey \thanks{e-mail:dbailey@ucar.edu} %
\and Tiffany Fourment \thanks{e-mail:tiffanyf@ucar.edu}
\and Becca Hatheway \thanks{e-mail:hatheway@ucar.edu}
\and Marika M. Holland \thanks{e-mail:mholland@ucar.edu}
\and Matt Rehme \thanks{e-mail:mrehme@ucar.edu}}
\affiliation{\scriptsize National Center for Atmospheric Research (NCAR) \\ University Corporation for Atmospheric Research (UCAR)}

\abstract{Data visualizations are ubiquitous in all disciplines and have become the primary means of analysing data and communicating insights. However, the predominant reliance on visual encoding of data continues to create accessibility barriers for people who are blind/vision impaired resulting in their under representation in Science, Technology, Engineering and Mathematics (STEM) disciplines.  This research study seeks to understand the experiences of professionals who are blind/vision impaired in one such STEM discipline (geosciences) in accessing data visualizations. In-depth, semi-structured interviews with seven professionals were conducted to examine the accessibility barriers and areas for improvement to inform accessibility research pertaining to data visualizations through a socio-technical lens. A reflexive thematic analysis revealed the negative impact of visualizations in influencing their career path, lack of data exploration tools for research, barriers in accessing works of peers and mismatched pace of visualization and accessibility research.  The article also includes recommendations from the participants to address some of these accessibility barriers.      
} 

\CCScatlist{
  \CCScatTwelve{Human-centered computing}{Visualization}{Visualization Design and evaluation methods}{};
  \CCScatTwelve{Human-centered computing}{Accessibility}{Accessibility technologies}{}
}

\begin{document}


\firstsection{Introduction}
\maketitle
 The immense popularity and pervasiveness of data visualizations have made them the defacto tools for data driven communication. However, this sole reliance on visual encoding for data interpretation creates significant accessibility barriers for people who are blind/vision impaired \cite{marriott_inclusive_2021, joyner_visualization_2022}. The impacts of inequities in information access are particularly amplified during times of crisis, as evidenced during COVID-19 \cite{siu_covid-19_2021} emphasizing the urgent need to improve the accessibility of data visualizations. Overall, accessibility considerations in data visualizations are an under-explored area in visualization research \cite{marriott_inclusive_2021}. Prior work has emphasized the need for visualization practitioners, accessibility researchers and people from the blind/low vision community to work together to examine the experiences of people with vision impairments as well as establish evidence-based guidelines to tackle the accessibility barriers \cite{marriott_inclusive_2021}.

The current research seeks to understand the experiences of professionals who are blind/vision impaired with accessing information presented through data visualizations in geosciences, one of the least diverse STEM disciplines\cite{carabajal_synthesis_2017}.  We conducted semi-structured, in-depth interviews with professionals who are blind/vision impaired, exploring the visualization accessibility barriers at various stages of their career.    

The following are the main contributions of this work:
\begin{itemize}
    \item Providing a descriptive account of the lived experiences of geoscience professionals who are blind/vision impaired
    \item Identifying social and technical barriers to visualization accessibility in geosciences
    \item Providing recommendations for addressing some of the accessibility barriers identified in the study and opportunities for accessible technology.
\end{itemize}

This study was designed to focus on the experiences of professionals from a single field of study (geosciences) in order to isolate any domain specific visualization considerations. However, this article covers the findings which could be broadly applied to other STEM disciplines.

\section{Related Work and Rationale}

The predominant focus of the research community on effective visual encoding of information inadvertently disenfranchises people with vision disabilities \cite{lee_reaching_2020,kim_accessible_2021}. This limited access to graphical material can have a snowball effect of exacerbating the overall under-representation of people with disabilities in broad disciplines. For instance, a previous study by Butler et al. \cite{butler_understanding_2017} has identified that people with vision impairments deliberately chose fields of study with fewer graphics, driven largely by the accessibility limitations with graphical material. These accessibility barriers in turn significantly reduced the career options and led to the disengagement of this population with their field of study, contributing to the negative perceptions of people in the field about the capacity of people with vision impairments \cite{atchison_professionally_2016,butler_understanding_2017}. Geosciences is one such STEM discipline which has been identified as one of the least diverse, with ~75\% fewer individuals with disabilities represented in the workforce than in the general population \cite{carabajal_synthesis_2017}.

A study by Atchinson and Libarkin \cite{atchison_professionally_2016} investigating the perceptions of geoscience professionals about accessibility found that practitioners perceived geosciences through a vision-centric lens, thereby assuming people with visual impairments to be least viable for geoscience careers among other types of disabilities. These perceptions of ability are reinforced through promotional materials and messaging (visualizations being one of them) that may be actively discouraging individuals with disabilities from considering geoscience as a viable career option \cite{atchison_professionally_2016}. To this end, providing access to information and programs was quoted to be one of the imperatives for successful inclusion of individuals with visual impairments in atmospheric science careers \cite{durre_untapped_2008}. Lastly, although this discussion has largely focused on the importance of accessibility for people with vision impairments, non-visual techniques of data representation have been known to be beneficial for people with intellectual and developmental disabilities \cite{wu_understanding_2021} as well as students without disabilities \cite{asher_teaching_2001} highlighting the broader impacts of inclusive design.   

Disability studies is an interdisciplinary field focused on the study of disability through social, cultural and political perspectives. The field has played a significant role in defining the rhetoric, language and methods through which to understand the experiences of people with disabilities \cite{mankoff_disability_2010}. A large portion of accessibility research articles in the visualization domain, published over the last two decades, focused on the evaluation of technologies and approaches to address specific accessibility barriers of data visualizations through technological interventions \cite{kim_accessible_2021}. This approach to accessibility, with a focus on actionable challenges and opportunities to address disability through technological interventions, draws its insights from the medical modal of disability \cite{mankoff_disability_2010}. In contrast, the social model defines disability as a consequence of exclusion brought forth by structural and environmental factors \cite{lundgard_accessible_2022}. Both these perspectives could help inform research and new areas of inquiry.  

Drawing from these concepts, the participatory design process for the creation of accessible content is an approach to design which engages people with disabilities (stakeholders) as equal participants throughout the design process from the outset (i.e, need-finding and problem definition stage) \cite{mankoff_disability_2010,lundgard_sociotechnical_2019}. This approach emphasises the lived experiences of stakeholders and reduces the likelihood of developing "disability dongles", a term coined by Liz Jackson to describe a "well intended elegant yet useless solution to a problem we never knew we had" \cite{lundgard_sociotechnical_2019}.  In their call to action for inclusive data visualizations, Marriot et al. \cite{marriott_inclusive_2021} identify the topic of understanding how people with disabilities currently access data and use data visualizations to be essential and largely missing from literature. They also identify that evidence based guidelines to facilitate the creation of accessible content is lacking, even among the data visualization practitioners who recognize the importance of accessibility.  

The current research study seeks to address the aforementioned shortcoming by conducting interviews with professionals who are blind/vision impaired, to understand their experiences accessing data visualizations in geosciences. While there have been a handful of studies exploring this topic \cite{butler_understanding_2017,sharif_understanding_2021,potluri_examining_2021}, the current work significantly differs from previous works in that we investigate this topic with a socio-technical lens, focus on a specific STEM discipline (geosciences) and do not limit the scope to any one particular visualization technique.

\section{Study Design}
We conducted in-depth, semi-structured interviews with professionals in geosciences who are blind/vision impaired exploring the research question: What are the experiences of students and professionals who are blind/vision impaired in geosciences in accessing information presented through data visualizations? 

\subsection{Participants}
The seven interviewees were individuals in geosciences who have experience with data visualizations either in their current work or through their education. They were recruited through professional contacts, snowball sampling and advertisements through geoscience listservs. The participants were offered a \$50 honorarium for their participation in a 50-60 minute interview. All the participants are legally blind and four of them were blind since birth. Four of the participants identified themselves as female and three as male. The participants' ages ranged between 25 and 65 years. All the participants assessed themselves to be highly proficient with computers. Five of the participants had a graduate degree and two of them had a highest level of bachelor's degree. Six of the participants currently worked in geosciences.

\subsection{Data Collection}
The interviews were conducted using Zoom video conferencing between June 2021 - March 2022. The interview questions were shared with the participants prior to the interview. The interview started with basic  questions followed by open-ended questions regarding participants' current occupation and technologies used to access visual graphics. This was followed by open-ended questions investigating three stages of their career: (i) K-12, (ii) Higher Education, and (iii) Professional. The questions explored the importance of data visualizations at each stage of their career, the impact of visual material on their career path, and the techniques used to address any barriers that they have encountered. They were also asked about their favorite visualization techniques and the reasons for that, as well as areas in visualization accessibility research that they felt needed to be addressed (Note: The interview guide is provided as a supplemental material accompanying this article.) The interviews were audio recorded and transcribed for data analysis. After each interview, preliminary analysis was conducted to gather insights and check the relevance of the data in addressing the research questions. Thematic saturation \cite{saunders_saturation_2018} was observed by the sixth interview.

\subsection{Data Analysis}
\subsubsection{Theoretical Assumptions}
The primary objective of the study was to draw insights on data visualization accessibility from the participants' experiences. This requires both the faithful reflection on the participant's narrative as well as the researcher's interpretive analysis to identify themes relevant to the research question. Consequently, reflexive thematic analysis (RTA) \cite{braun_successful_2013,braun_reflecting_2019,braun_one_2021} was determined to be the best approach for data analysis. This study applied a constructionist epistemology with an experiential orientation. This approach acknowledges the importance of both recurrence and meaningfulness of a communicated experience while enabling the researcher to preserve the subjectivity in a participant's experience\cite{byrne_worked_2021}. Given the data-driven, exploratory nature of this study, a predominantly inductive approach was adopted for the analysis. Lastly, in order to focus on the experiences and the reality of the participants, semantic coding (descriptive analysis) was prioritized over latent coding (interpretive analysis) whenever possible acknowledging that a certain degree of latent coding was necessary in order to identify themes relevant to the research questions.

\subsubsection{Analytical Process}
The six phase, non-sequential, recursive process for RTA proposed by Braun and Clarke \cite{braun_successful_2013,byrne_worked_2021} was followed to analyse the transcribed interviews. Following the familiarization phase, descriptive and interpretive open-codes were generated for each piece of information that were relevant to the research questions. Afterwards, these initial codes were reviewed and patterns of codes were examined for prospective themes using affinity diagramming. This recursive process was applied for the first two interviews and a list of preliminary candidate themes was produced. The remaining interviews were then coded by comparing with the existing candidate themes, reviewing both the relevance of the new codes and the themes in answering the research questions. The candidate themes thus produced were then reviewed following Patton's dual criteria for judging categories \cite{patton_qualitative_1990,byrne_worked_2021} and a final set of themes and sub-themes were generated. Following the recommendations of Nowell et al. \cite{nowell_thematic_2017} validity of this approach was determined by maintaining a clear audit trail, peer debriefing, transferability documented through thick descriptions (although only a summary is provided in this article).

\section{Results and Discussion}
The first iteration resulted in around 75 codes which were later distilled down to 23 unique codes by merging similar codes and removing duplicates. These codes were organized into 5 key findings and 3 recommendations.

\subsection{Key Findings}
\subsubsection{A deterrent}
A common observation that came up repeatedly in all the interviews was that visualization accessibility continues to be a significant hindrance. While it did not stop them from continuing in the field, it posed a significant challenge in their careers. Participants made statements such as:
\begin{displayquote}
\textit{"I don't want 
to say it's career changing, because I didn't change my career, but 
it's been career challenging." - P1
} 
\end{displayquote}

\begin{displayquote}
\textit{"I do feel like I'm 
missing out on a lot. In not like looking at, like, the data 
visualizations, or understanding them." - P6
}
\end{displayquote}

This theme revealed that accessibility limitations of visualizations inadvertently served as a deterrent for people with vision impairment from pursuing a career in geosciences, potentially contributing to the training barriers to inclusion identified in a previous study \cite{atchison_professionally_2016}.  

\subsubsection{Limited tools for research}  Some of common visualization techniques used in geosciences are line charts for time series data, contour plots and geographic heat maps for 2D scalar data (E.g. Temperature, Pressure) and, quiver plots for vector fields (E.g. wind, vorticity). Participants also mentioned coming across occasional 3D visualizations such as iso-surfaces. These visualizations are used both in research as well as education and outreach. While there have been significant technological advancements enabling the use of alternate modalities for data exploration such as tactile graphics, participants found that these approaches had limitations which prevented their application in research. The common limitations mentioned were: inability of the existing techniques to accurately encode the information at a \textbf{resolution} needed for research, lack of \textbf{refreshability}, slow \textbf{production time}, high \textbf{cost} and lack of on-demand \textbf{autonomous} techniques for independent exploration.  Consequently, participants stated that seeking assistance from a sighted person with subject matter knowledge was currently the most effective means to access data visualizations. This reliance on sighted assistance limited their ability to participate in a scientific discourse as two participants stated:

\begin{displayquote}
\textit{"One 
of the workarounds for this was collaboration and that worked 
well. But nonetheless, I was constrained in what I could independently 
do to duplicate data." - P3}
\end{displayquote}

\begin{displayquote}
\textit{"I was completely reliant on 
them, to tell me what their data said, Now, that's actually useful. 
It's good to have it if you want the student to tell you what the data 
said. But you also want to be able to check that independently." - P2}
\end{displayquote}

One approach to independent data exploration that participants found to be useful and promising was sonification, where data is encoded in audible frequencies \cite{hermann_sonification_2011}. However, their application was limited to time series plots and access to multi-dimensional data, vector fields and comparative visualizations remain an ongoing challenge.


\subsubsection{Accessibility barriers in published work} This theme captures the barriers identified by the participants in accessing the visualizations published by peers in presentations, posters and journal articles.  Although there exist guidelines \cite{noauthor_web_2018} for including alternate textual descriptions (alt text) in images, participants mentioned that a majority of visualizations in academic works either lack alt text or have inadequate description for them to comprehend the visuals being presented. This puts the onus on the person with the disability to reach out to the presenter/author, which might not always be feasible and creates a barrier for their participation in the scientific discourse. Participants made statements such as:       

\begin{displayquote}
\textit{"Don't even get me started on conferences ... conferences are the places where I have the most difficulty with data visualizations." - P4}
\end{displayquote}

\begin{displayquote}
\textit{"I'm 
on a crusade right now here to help this institution make its all its 
documents more accessible, including Excel, mostly in the office 
suite, Excel, PowerPoint, images and PowerPoint with alt text." - P1}
\end{displayquote}

\begin{displayquote}
\textit{"Some of those constraints about looking at other people's data were also quite serious. So things like seminars, and so on were and are still problematic." -P2 }
\end{displayquote}

Lundgard et al. \cite{lundgard_sociotechnical_2019} mention the awareness of accessibility guidelines as one of the considerations for accessible visualizations design. This theme revealed the lack of awareness in academia not only regarding the guidelines but also the commonly adopted best practices for graphics accessibility adopted by industry such as alt text.    

\subsubsection{Degree of impact vs career stage}
 Participants who were in their late career mentioned that they received sighted assistance through staff employed by their institution or funded through research grants both of which are not readily available for graduate students and early career professionals. The same participants also mentioned that data analysis was a significant part of job duties for people in their early career stages. Although the concept of data visualizations was introduced in K-12 education, participants were expected to produce visualizations in their graduate school.  While they managed to fulfil course requirements through alternate accommodations and creative ways to demonstrate their understanding of core concepts, the lack of tools for independent data exploration implicitly influenced their subsequent career choices. Participants made statements such as:    

\begin{displayquote}
\textit{"I was being nudged into areas where this [visualization] was less 
of a limitation. So the more theoretical, and the more 
computational
... It [visualization] was a constraint. And I would 
have had a wider choice of topics and styles of accessing data and 
working with data had it been available.
" - P2}
\end{displayquote}

\begin{displayquote}
\textit{"The lack of data visualization tools definitely made it harder. It influenced the field I
chose in the end." - P5}
\end{displayquote}

Butler et al. \cite{butler_understanding_2017} have investigated the state of accessible graphics in higher education and have identified significant issues impacting the enrollment and retention of students with vision impairments in STEM fields.  The current theme has identified similar experiences faced by students in geosciences and highlights the impact of the higher education experience in determining the retention of students in the field. 

\subsubsection{Technological advancements: a blessing or a barrier?}
Reflecting on the evolution of tools and techniques to access visualizations, the participants ascertained the overall positive impact of technological advancements in improving the accessibility barriers with statements such as:
\begin{displayquote}
\textit{"As far as data visualization goes, they've [technological advancements] helped immeasurably. This interview was actually quite useful to 
make me think about data access now versus say, my early graduate 
school, and the case has immeasurably improved." -P2
}
\end{displayquote}
However, the broader technological advancements have also contributed to the increasing complexity of visualization tools and techniques, which exacerbated accessibility problems as one participant stated:     

\begin{displayquote}
\textit{"Technology has enabled more flashy data visualizations, which the scientific community
has increasingly relied on. This has caused a migration of papers and presentations to
become more visual, which leaves a lot of people out. -P6"
}
\end{displayquote}

This theme revealed that accessibility considerations have not received equivalent attention compared to mainstream visualization techniques. This mismatched pace could further widen the accessibility gap and highlights the importance and urgency of accessibility research in data visualizations.
\subsection{Recommendations} 
In addition to the above themes, three recommendations emerged from the interviews regarding strategies to address some of the barriers identified by the participants:  

\paragraph{\textbf{Raise accessibility awareness.}}  The first recommendation addresses the absence of basic accessibility features such as alternate textual descriptions in most of the works published in academia which was mentioned by almost all the participants who used screen-readers. It also emphasises the importance of understanding the accessibility guidelines and adopting the current best practices to improve the accessibility of graphics in academic works.  Further, given that most of the recommendations for alternate textual descriptions do not necessarily translate to scientific visualizations, it is important to determine the most effective way to describe data visualizations by consolidating the existing guidelines from different sources and/or building on them.        

\paragraph{\textbf{Provide access to the underlying data.}}  Many participants in this study devised their own software and techniques to interpret data. Some of the most commonly used approaches, such as the use of a braille refreshable display to scan tabular data and code snippets to sonify time-series data, are only usable if they are able to access the underlying data used to generate the visualizations.  To this end, participants identified that providing access to the underlying data would be a significant step towards improving the accessibility of visualization requiring minimal effort from the presenter.   

\paragraph{\textbf{Design inclusive technologies with a multi-sensory approach.}} Visualizations are without doubt one of the most effective means of communicating information.  The main limitation of visualizations is their predominant reliance on a single sensory modality.  This last recommendation acknowledges human diversity and encourages the consideration of more than one sensory modality in representing data and information. Even among people with disabilities, the wide spectrum of vision disabilities necessitates solutions that do not constrain individuals to use a single approach to interaction.

\section{Summary}
We examined the experiences of geoscience professionals who are blind/vision impaired with accessing data visualizations, using reflexive thematic analysis. The study revealed that visualization access continues to be a significant barrier in their careers. The impact was more pronounced for students and early career professionals and these barriers often influenced their career decisions and limited their options. The study also revealed social barriers such as difficulties in accessing visualizations appearing in academic works due to the lack of awareness among peers regarding the common best practices for making graphics accessible; and technical barriers such as lack of accessible data exploration tools for research and underscored the mismatched pace of visualization and accessibility research. Lastly, recommendations were provided by the participants highlighting the importance of both technological development and attitudinal shifts needed to address the accessibility barriers. 
Although the focus of this study was on professionals from geosciences, the findings reported in this article are more generally applicable to any other field and reinforce some of the observations reported in previous studies \cite{joyner_visualization_2022,lundgard_sociotechnical_2019,marriott_inclusive_2021}.  

Additional areas that are worth exploring in future studies based on the data and insights gathered are listed below.
\begin{itemize}
    \item Visualizations have both exploratory (visualizations used in research) and explanatory (visualizations used for communication and story telling) applications \cite{iliinsky_designing_2011}. Some of the technological limitations mentioned by the participants seemed to depend heavily on the purpose of visualization. Any study investigating accessibility considerations with data visualizations need to factor in this distinction.  
    \item There also seems to be a difference between techniques used to absorb information from data visualizations and, techniques used to create visualizations to communicate with others. There are very few content creation tools that are accessible for people who are blind/vision impaired and this needs to be investigated further for identifying areas of improvement.  
    \item 3D data exploration continues to be a major limitation for people who are blind/vision impaired. Multi-modal visualization techniques using haptics in immersive environments such as augmented, virtual and mixed reality were mentioned by one of the participant to be promising for exploring 3D data, based on their experience with a research prototype. User-interface design in immersive environments is a rapidly growing area of research. Future studies exploring the accessibility applications of these emerging technologies in data visualizations could help address the gaps identified in this and other related works.
\end{itemize}

We hope that this work helps motivate both researchers and professionals in wider STEM disciplines to develop more inclusive data visualizations.

\acknowledgments{
We would like to thank the participants for their time in sharing their experiences and recommendations. We would also like to thank Dr. Bhoomika Bhagchandani for her valuable guidance on the study design and thematic analysis. We sincerely appreciate the feedback received from the anonymous reviewers who helped us improve this article. This work was supported in part by the 2020 University Corporation For Atmospheric Research President's Strategic Initiative Fund and the National Center for Atmospheric Research, which is a major facility sponsored by the National Science Foundation under Cooperative Agreement 1852977.}

\bibliographystyle{abbrv-doi}

\bibliography{references}
\end{document}